# HOW MANY MUONS DO WE NEED TO STORE IN A RING FOR NEUTRINO CROSS-SECTION MEASUREMENTS?

Steve Geer

Fermi National Accelerator Laboratory, PO Box 500, Batavia, IL 60510

ABSTRACT

Analytical estimate of the number of muons that must decay in the straight section of a storage ring to produce a neutrino & anti-neutrino beam of sufficient intensity to facilitate cross-section measurements with a statistical precision of 1%.

## 1. INTRODUCTION

As we move into the era of precision long-baseline $\nu_\mu \to \nu_e$ and $\bar{\nu}_\mu \to \bar{\nu}_e$ measurements there is a growing need to precisely determine the $\nu_e$ and $\bar{\nu}_e$ cross-sections in the relevant energy range, from a fraction of 1 GeV to a few GeV. This will require $\nu_e$ and $\bar{\nu}_e$ beams with precisely known fluxes and spectra. One way to produce these beams is to use a storage ring with long straight sections in which muon decays ($\mu^- \to e^- \nu_\mu \bar{\nu}_e$ if negative muons are stored, and $\mu^+ \to e^+ \nu_e \bar{\nu}_\mu$ if positive muons are stored) produce the desired beam. The challenge is to capture enough muons in the ring to obtain useful neutrino and anti-neutrino fluxes. Early proposals to use a muon storage ring for neutrino oscillation experiments were based upon injecting "high energy" charged pions into the ring which then decayed to create stored muons. These proposals were hampered by lack of sufficient intensity to pursue the physics. The Neutrino Factory proposal in 1997 was designed to fix this problem by using a Muon Collider class "low energy" muon source to capture many more pions at low energy, allow them to decay in an external decay channel, manipulate their phase space to capture as many muons as possible within the acceptance of an accelerator, and then accelerate to the energy of choice before injecting into a specially designed ring with long straight sections. All this technology would do a wonderful job in fixing the intensity problem, but at a price that excludes this solution from being realized in the short term. The question that we are now faced with is whether the older, lower intensity "parasitic" muon storage ring based on "high energy" pion decays can, with suitable modification, produce sufficient intensity to measure the desired cross-sections. Fortunately, the intensity requirements for cross-section measurements are less demanding than the corresponding requirements for oscillation measurements, so there is hope. To fuel the discussion, in this note we consider the design goal: how many muons do we need to store?

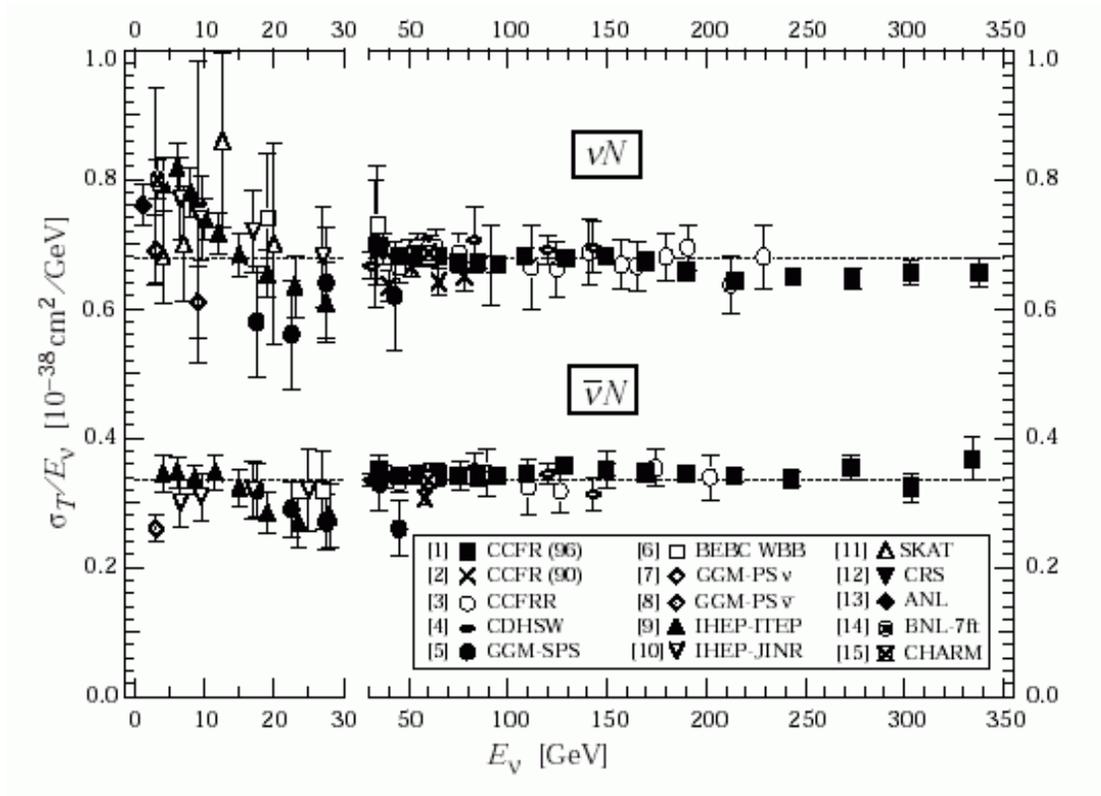

FIGURE 1: Muon neutrino and anti-neutrino CC cross-sections. Figure from K. Hagiwara et al., Phys. Rev. D 66 (Review of Particle Physics 2002).

## 2. EVENT RATES

Muon decay kinematics is precisely known. The spectra of the daughter neutrinos and anti-neutrinos are described by simple analytical expressions. In the muon rest-frame the distribution of energies and angles is given by:

$$\frac{d^2 N_{\nu_\mu}}{dx d\Omega_{c.m.}} \propto \frac{2x^2}{4\pi}\left[(3-2x)+(1-2x)P_\mu \cos\theta_{c.m.}\right],$$

$$\frac{d^2 N_{\bar{\nu}_e}}{dx d\Omega_{c.m.}} \propto \frac{12x^2}{4\pi}\left[(1-x)+(1-x)P_\mu \cos\theta_{c.m.}\right],$$

where $x \equiv 2E_\nu/m_\mu$, $\theta_{c.m.}$ is the angle between the neutrino momentum vector and muon spin direction, and $P_\mu$ is the average muon polarization along the beam direction. The corresponding $\bar{\nu}_\mu$ and $\nu_e$ distributions for $\mu^+$ decay are obtained by changing $P_\mu \to -P_\mu$. In the forward direction ($\cos\theta_{lab} \approx 1$) the maximum $E_\nu$ in the laboratory frame $E_{max} = \gamma[1+\beta\cos\theta_{c.m.}]m_\mu/2$, and:

$$\frac{d^2N_{\nu_\mu}}{dxd\Omega_{lab}} \propto \frac{1}{\gamma^2(1-\beta\cos\theta_{lab})^2} \frac{2x^2}{4\pi}\left[(3-2x)+(1-2x)P_\mu\cos\theta_{c.m.}\right],$$

$$\frac{d^2N_{\bar{\nu}_e}}{dxd\Omega_{lab}} \propto \frac{1}{\gamma^2(1-\beta\cos\theta_{lab})^2} \frac{12x^2}{4\pi}\left[(1-x)+(1-x)P_\mu\cos\theta_{c.m.}\right].$$

Note that in principle, if the muons are polarized, then manipulating $P_\mu$ can be used to modify the neutrino and anti-neutrino spectra. The spectrum of interactions in an experiment is given by the convolution of these distributions with the energy dependent cross-sections. If $E_\nu >$ ~10 GeV the cross-sections are dominated by deep inelastic scattering and proportional to $E_\nu$:

$\sigma(\nu + N \to l^- + X) \approx 0.67 \times 10^{-38}$ cm$^2 \times E_\nu$ (GeV)

$\sigma(\bar{\nu} + N \to l^+ + X) \approx 0.34 \times 10^{-38}$ cm$^2 \times E_\nu$ (GeV)

At lower energies it is more complicated. However, in the following we use a linear dependence down to lower energies to get a first estimate of event rates. The data in Figure 1 suggest this is not a bad first approximation. The number of neutrino and anti-neutrino charged current (CC) events per incident neutrino in an isoscalar target is given by:

N ($\nu + N \to l^- + X$) $\approx 4 \times 10^{-15} \times E_\nu$ (GeV) events per gm/cm$^2$

$$N(\bar{\nu} + N \to l^+ + X) \approx 2 \times 10^{-15} \times E_\nu \text{ (GeV)} \text{ events per gm/cm}^2$$

Integrating over the energy distribution, the event rates (events per kt per muon-decay) are given by:

$$N(\nu_e + N \to l^- + X) \approx 1.2 \times 10^{-14} \times [(E_\mu / \text{GeV})^3 / (L / \text{km})^2] \times 0.6 \times (1 - P_\mu)$$

$$N(\nu_\mu + N \to l^- + X) \approx 1.2 \times 10^{-14} \times [(E_\mu / \text{GeV})^3 / (L / \text{km})^2] \times (0.7 + 0.3 P_\mu)$$

$$N(\bar{\nu}_e + N \to l^+ + X) \approx 0.6 \times 10^{-14} \times [(E_\mu / \text{GeV})^3 / (L / \text{km})^2] \times 0.6 \times (1 + P_\mu)$$

$$N(\bar{\nu}_\mu + N \to l^+ + X) \approx 0.6 \times 10^{-14} \times [(E_\mu / \text{GeV})^3 / (L / \text{km})^2] \times (0.7 - 0.3 P_\mu)$$

where L is the distance from the decaying muon to the detector. In practice, the length of the beam forming straight section in the muon storage ring ($L_S$) cannot be neglected, and may even be comparable to the distance from the end of the straight section to the near detector ($L_D$). We must therefore replace $1/L^2$ in the event rate expressions with $1/[L_D(L_D+L_S)]$, yielding:

$$N(\nu_e + N \to l^- + X) \approx 1.2 \times 10^{-14} \times [(E_\mu / \text{GeV})^3 / (L_D^2 + L_D L_S)] \times 0.6 \times (1 - P_\mu)$$

$$N(\nu_\mu + N \to l^- + X) \approx 1.2 \times 10^{-14} \times [(E_\mu / \text{GeV})^3 / (L_D^2 + L_D L_S)] \times (0.7 + 0.3 P_\mu)$$

$$N(\bar{\nu}_e + N \to l^+ + X) \approx 0.6 \times 10^{-14} \times [(E_\mu / \text{GeV})^3 / (L_D^2 + L_D L_S)] \times 0.6 \times (1 + P_\mu)$$

$$N(\bar{\nu}_\mu + N \to l^+ + X) \approx 0.6 \times 10^{-14} \times [(E_\mu / \text{GeV})^3 / (L_D^2 + L_D L_S)] \times (0.7 - 0.3 P_\mu)$$

events per kt per muon-decay, where all distances are in km. If we are interested in the rates per muon stored (rather than per muon decay in the straight section) we must multiply the above by $0.5 L_S / (L_S + L_A)$, where $L_A$ is the length of an arc.

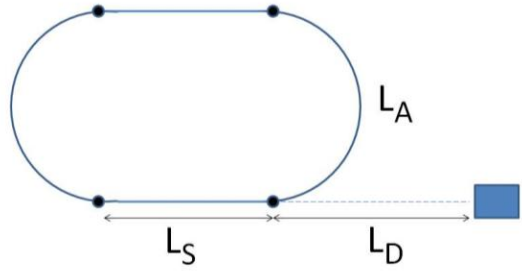

FIGURE 2: Ring geometry.

## 3. OPTIMIZING THE GEOMETRY

To maximize the flux we want (i) the arc lengths $L_A$ to be as short as practical consistent with using realistic bending magnets, and (ii) $L_D$ to be as short as possible whilst providing sufficient shielding for the detector. The optimum choice for the straight section length $L_S$ depends upon whether the muons are created externally and then injected into the storage ring, or are created in situ by charged pions decaying in the injection straight section.

### 3.1 Muons Created Externally

In this case there is an optimum straight section length that maximizes the flux at the detector. The fraction of muons that decay in the beam forming straight section: $F \approx 0.5 \times L_S / (L_S + L_A)$. The flux $\Phi_\nu$ in a detector at a distance $L_D$ from the downstream end of the straight section

$\Phi_\nu \propto L_S / [\,(L_S + L_A) \times L_D \times (L_S + L_D)\,]$.

The flux can be maximized by choosing $L_S = \sqrt{(L_A L_D)}$.

### 3.2 Muons Created by Pion Decays within the Straight Section

If the muons are produced by charged pions decaying within the injection straight section, then a larger $L_S$ will enhance the fraction of pions that decay to produce captured muons, favoring a straight section with $L_S > \sqrt{(L_A L_D)}$. In practice, for pions with energies O(1) GeV or more, the pion decay length $\gamma c \tau \gg L_S$, and to a first approximation the number of useful pion decays increases linearly with $L_S$. Hence, the flux $\Phi_\nu$ in a detector at a distance $L_D$ from the downstream end of the straight section is proportional to the purely geometric factor $F_G$:

$$\Phi_\nu \propto F_G \equiv 20\, L_S^2 / [\,(L_S + L_A) \times L_D \times (L_S + L_D)\,]$$

where all distances are in meters and the normalization is fixed so that for $L_D = 20$m, $F_G \to 1$ as $L_S \to \infty$. The geometrical factors are listed in Table 1 for some representative cases. Note that to maximize the neutrino flux we would like $\gamma c \tau \gg L_S \gg L_A, L_D$.

**TABLE 1:** Geometrical factors $F_G$ shown as a function of the distance from the downstream end of the straight section to the detector $L_D$, the arc length $L_A$, and the straight section length $L_S$. Note that in the first 2 rows $L_S = \sqrt{(L_A L_D)}$, in the second 2 rows $L_S = 2\sqrt{(L_A L_D)}$, and in the last 2 rows $L_S$ relatively large.

.

| $L_D$ (m) | $L_A$ (m) | $L_S$ (m) | $F_G$ |
|---|---|---|---|
| 20 | 20 | 20 | 0.25 |
| 20 | 40 | 28.284 | 0.24 |
| 20 | 20 | 40 | 0.44 |
| 20 | 40 | 56.569 | 0.43 |
| 20 | 20 | 100 | 0.69 |
| 20 | 40 | 100 | 0.6 |

## 4. AN EXAMPLE

As an example, consider a 3 GeV unpolarized muon beam ($P_\mu = 0$) stored in a ring with arc length $L_A = 36$m, and distance to the detector $L_D = 20$m. If the muons are created externally, the optimum straight section length $L_S = 27$m. The fraction of the muons that decay in the beam-forming straight-section is then 0.21. The event rates, per stored muon are listed in Table 2 for some chosen $L_A$ (based on a preliminary lattice design by M. Popovic). Rates are shown for a 1 ton detector as a function of the stored muon energy.

**TABLE 2:** CC neutrino and anti-neutrino event rates per stored muon for an unpolarized beam stored in a ring with arcs of length $L_A$, and optimized straight section length $L_S$ (assuming $L_D = 20$m and the muons are produced in an external decay channel). The rates are per stored muon for a 1 ton detector.

| $E_\mu$ GeV | $L_A$ km | $L_S$ km | $L_D$ km | $\nu_e$ events | $\nu_e$ bar events | $\nu_\mu$ events | $\nu_\mu$ bar events |
|---|---|---|---|---|---|---|---|
| 1 | 0.012 | 0.0155 | 0.02 | 2.9E-15 | 1.4E-15 | 3.3E-15 | 3.3E-15 |
| 2 | 0.024 | 0.0219 | 0.02 | 1.6E-14 | 8.2E-15 | 1.9E-14 | 1.9E-14 |
| 3 | 0.036 | 0.0268 | 0.02 | 4.4E-14 | 2.2E-14 | 5.2E-14 | 5.2E-14 |
| 4 | 0.048 | 0.031 | 0.02 | 8.9E-14 | 4.4E-14 | 1E-13 | 1E-13 |
| 5 | 0.06 | 0.0346 | 0.02 | 1.5E-13 | 7.5E-14 | 1.8E-13 | 1.8E-13 |

## 3. CONCLUSION

If we want to make $\nu_e$ and anti-$\nu_e$ cross section measurements in the 0.5 GeV – 3 GeV range, for example, and wish to achieve a statistical precision of O(1%), we will need $O(10^4)$ events per measurement. The analytical estimates above imply that we will need an "exposure" that depends on the storage ring energy, and for a 3 GeV ring is $O(10^{18})$ stored muons × tons per measurement. An order of magnitude more muons would be required for correspondingly precise very low energy cross-section measurements with a 1 GeV storage ring.

## 5. OTHER ESTIMATES

Numerical calculations which are consistent with the analytical results in this note can be found in Campanelli, Navas-Concha and Rubbia, [arXiv:hep-ph/0107221v1](arXiv:hep-ph/0107221v1). In addition, event rates using the existing antiproton Debuncher are given in S. Geer, FERMILAB-FN-706 (2001), which uses estimates based on numbers scaled from the earlier P-860 proposal (M.J. Murtagh et al.).

*ACKNOWLEDGEMENT*

This work is supported at the Fermi National Accelerator Laboratory, which is operated by the Fermi Research Association, under contract No. DE-AC02-76CH03000 with the U.S. Department of Energy.